\documentclass[twocolumn,amssymb, nobibnotes, aps, prd, nofootinbib]{revtex4-2}

\usepackage{amsmath}
\usepackage{amsfonts}
\usepackage{amssymb}
\usepackage{booktabs}
\usepackage[T1]{fontenc}
\usepackage{graphicx}
\usepackage[colorlinks=true, linkcolor=darkgray, urlcolor=darkgray, citecolor=darkgray]{hyperref}
\usepackage{siunitx}
\usepackage{xcolor}

\usepackage{graphicx}	
\usepackage{amsmath}	
\usepackage{amssymb}	
\usepackage{siunitx}
\usepackage{xcolor}

\usepackage{newtxtext,newtxmath}

\usepackage{tikz}
\usetikzlibrary{graphs}

\graphicspath{{figures/}}

\DeclareSIUnit\year{yr}
\DeclareSIUnit\parsec{pc}

\newcommand{\mywidth}{0.8\columnwidth}

\newcommand{\depth}{\mathcal{D}}

\newcommand{\ellip}{\varepsilon}
\newcommand{\ellipsd}{\ellip^{\mathrm{sd}}}
\newcommand{\ellipul}{\ellip^{\mathrm{ul}}}

\newcommand{\fB}{f_{\mathrm{B}}}

\newcommand{\fauld}{f_{\mathrm{max}}}

\newcommand{\hul}{h_0^{\mathrm{ul}}}

\newcommand{\KAbeta}{\beta}
\newcommand{\KAbetaLong}{\frac{5}{128\pi^4}\frac{c^5}{GI}}
\newcommand{\KAbetaFiducial}{
    \SI{1.455e11}{\per\second\cubed}
    \left(\frac{I}{\SI{1e38}{\kilogram \metre\squared}}\right)^{-1}
    }
\newcommand{\logellip}{\log_{10}{\ellip}}

\newcommand{\pellip}[2]{\logellip \in [#1, #2]}

\newcommand{\pdet}{p_{\mathrm{det}}}
\newcommand{\pt}{\tilde{t}}
\newcommand{\ptaug}{\tilde{\tau}_{\mathrm{g}}}

\newcommand{\sqrtSn}{\sqrt{S_{\mathrm{n}}}}
\newcommand{\sdown}[2]{f_{#1}^{\mathrm{#2}}}
\newcommand{\tage}{t_{\mathrm{age}}}

\newcommand{\taug}{\tau_{\mathrm{g}}}

\newcommand{\Tobs}{T_{\mathrm{obs}}}

\newcommand{\rhog}{\rho_{\mathrm{g}}}
\newcommand{\Ng}{N_{\mathrm{g}}}
\newcommand{\Vt}{V}

\begin{document}
\title{
    Blind-search constraints on the sub-kiloparsec population of continuous gravitational-wave sources
}

\author{Rodrigo Tenorio}
\email{rodrigo.tenorio@ligo.org}
\affiliation{
    Departament de F\'isica, Universitat de les Illes Balears, IAC3--IEEC,\\
    Cra.~de Valldemossa km 7.5, E-07122 Palma, Spain
    }

\begin{abstract}
    We use the latest all-sky continuous gravitational-wave (CW) searches to estimate constraints
    on the \mbox{sub-kiloparsec} population of unknown neutron stars (NS). We then extend this
    analysis to the forthcoming \mbox{LIGO-Virgo-KAGRA} observing runs and the third generation (3G) of 
    ground-based interferometric detectors (Einstein Telescope and Cosmic Explorer).
    We find that sources with ellipticities greater than $\ellip\gtrsim 10^{-7}$ can be well-constrained
    by current and future detectors regardless of their frequency.
    3G detectors will extend these constraints down to \mbox{$\ellip \gtrsim 10^{-8}$} across the
    whole sensitive band and \mbox{$\ellip \gtrsim 10^{-9}$} above \SI{1}{\kilo \hertz}. We do not
    expect \mbox{$\ellip \lesssim 10^{-8}$} sources to be constrained below \SI{1}{\kilo\hertz}.
    Finally, we discuss the potential impact of using astronomical priors on all-sky searches in terms
    of sensitivity and computing cost. The populations here described can be used as a guide to set up
    future all-sky CW searches.
\end{abstract}

\maketitle

\section{Introduction}
\label{sec:introduction}
Blind searches for continuous gravitational-wave signals (CWs) attempt
to detect long-lasting gravitational waves from
unknown rapidly-spinning neutron stars (NSs)~\cite{Tenorio:2021wmz, Riles:2022wwz}.
These signals may be emitted due a deformation on the
NS's crust or \mbox{r-mode} instabilities~\cite{Sieniawska:2019hmd},
and would fall within the sensitive band of ground-based interferometric detectors such as
Advanced LIGO~\cite{AdvancedLIGO},
Advanced Virgo~\cite{AdvancedVirgo}, 
KAGRA~\cite{KAGRA:2018plz}, and the planned third-generation detectors (3G)
Einstein Telescope (ET)~\cite{Maggiore:2019uih} and
Cosmic Explorer (CE)~\cite{Reitze:2019iox}.

All-sky searches are a subtype of blind searches which cover the whole sky,
a broad frequency band, and any extra parameters required by the signal model.
Due to their breadth, fully-coherent matched filtering 
becomes computationally unfeasible and semicoherent methods,
which require a much lower computing cost at the expense of sensitivity,
must be used~\cite{Krishnan:2004sv}. 

Semicoherent methods divide the data into segments which are individually
analyzed. This relaxes the signal model to allow for phase discontinuities
between consecutive segments, which increases the robustness of a search
to unaccounted physical phenomena,
such as accretion-induced spin-wandering~\cite{Mukherjee:2017qme} or NS 
glitches~\cite{Ashton:2017wui}. These properties make blind searches
an ideal tool to study the nearby population of unknown CW sources.

Population-synthesis analyses in~\cite{Popov:2003hq, Popov:2003iq} 
suggest that within \SI{600}{\parsec} there should be $100$ to $ 140$
NSs younger than $\SI{4}{\mega\year}$, mainly around the Gould belt~\cite{2014Ap.....57..583B}.
This is about a factor 3 higher than the current number listed in the
ATNF catalog~\cite{2005AJ....129.1993M, atnf_link}.
The analyses in~\cite{2010AN....331..349H, 2014AN....335..935S}
support the  Gould belt to be a region with an enhanced supernova rate with respect to the
Galactic average and suggest sub-kiloparsec unobserved young NSs are located  within 
$10\%$ of the sky.

In this paper, we use blind CW searches in the third observing run of
the LIGO-Virgo-KAGRA collaboration (O3)~\cite{KAGRA:2023pio} to estimate constraints
on the sub-kiloparsec population of young unknown NSs. 
We model these NSs as gravitars~\cite{Palomba:2005na}, which emit all their energy
as CWs and provide an optimistic detectability upper limit.
We then project these constraints into future observing runs (O4, O5) and 3G detectors.

The paper is structured as follows: in Sec.~\ref{sec:ellipticity_from_uls}, we review
the latest upper limits produced by three all-sky searches in LIGO-Virgo-KAGRA O3 data.
In Sec.~\ref{sec:gravitar_population}, we introduce and explore the properties of a 
population of young gravitars. In Sec.~\ref{sec:sensitivity}, we estimate the detectability 
of several populations of young gravitars depending on their ellipticity;
these results are then discussed in Sec.~\ref{sec:discussion}.
Conclusions are presented in Sec.~\ref{sec:conclusion}.

\section{Ellipticity upper limits from blind searches}
\label{sec:ellipticity_from_uls}
Rapidly-rotating NSs sustaining an equatorial ellipticity $\ellip$
are expected to emit CW signals with an amplitude given by~\cite{Jaranowski:1998qm}
\begin{equation}
    h_0(f_0, \ellip, d) = \frac{4 \pi^2 G}{c^4}I \frac{f_0^2 \ellip}{d} \,,
    \label{eq:h0}
\end{equation}
where $f_0$ is the CW frequency measured in \unit{Hz}
(twice the rotational frequency in this model) and 
$d$ is the distance from the NS to the detector.
$I=\SI{1e38}{\kilo\gram \,\meter^2}$ is the star's canonical moment of inertia
around the rotational axis. This quantity is uncertain by a factor of a
few~\cite{Owen:2005fn,Bejger:2005jy, Horowitz:2009vm,LIGOScientific:2007leh};
in this work, we assume the canonical value and operate in terms of ellipticity.

The main data product of CW searches are upper limits
on the minimum amplitude detectable by the search at a certain
confidence level $\hul$. Upon assuming an emission model, such as
Eq.~\eqref{eq:h0}, these upper limits can be translated into
upper bounds on other quantities, such as the maximum allowed ellipticity
on a source at a certain distance
\begin{equation}
    \ellipul(f_0, d) = \frac{c^4}{4 \pi^2 G}\frac{d}{I}\frac{\hul}{f_0^2} \,.
    \label{eq:ellip_ul}
\end{equation}
These constraints, however, do not hold for arbitrarily high ellipticities.
A given search covers a limited range of spindown values $f_1$.
The maximum ellipticity covered by a search, thus,
corresponds to that of a source spinning down at the maximum rate
solely due to CW emission~\cite{LIGOScientific:2007leh}
\begin{equation}
    \ellipsd(f_0, f_1)= \sqrt{
        \frac{5 c^5}{32 \pi^4 G I}
        \frac{-f_1}{f_0^{5}}
        }\;.
    \label{eq:ellipsd}
\end{equation}
For given distance $d$, and a given frequency $f_0$, a blind search returns
an exclusion range of ellipticities
\mbox{$\ellipul(f_0, d) \lesssim \ellip \lesssim \ellipsd(f_0, f_1)$}.
Note that $\ellipul$ depends on the distance to the source,
whereas $\ellipsd$ does \emph{not}; as a result, the spindown range
of a search sets a \emph{maximum distance} beyond  which sources fall 
outside the search's parameter space.

\begin{figure}
    \includegraphics[width=\columnwidth]{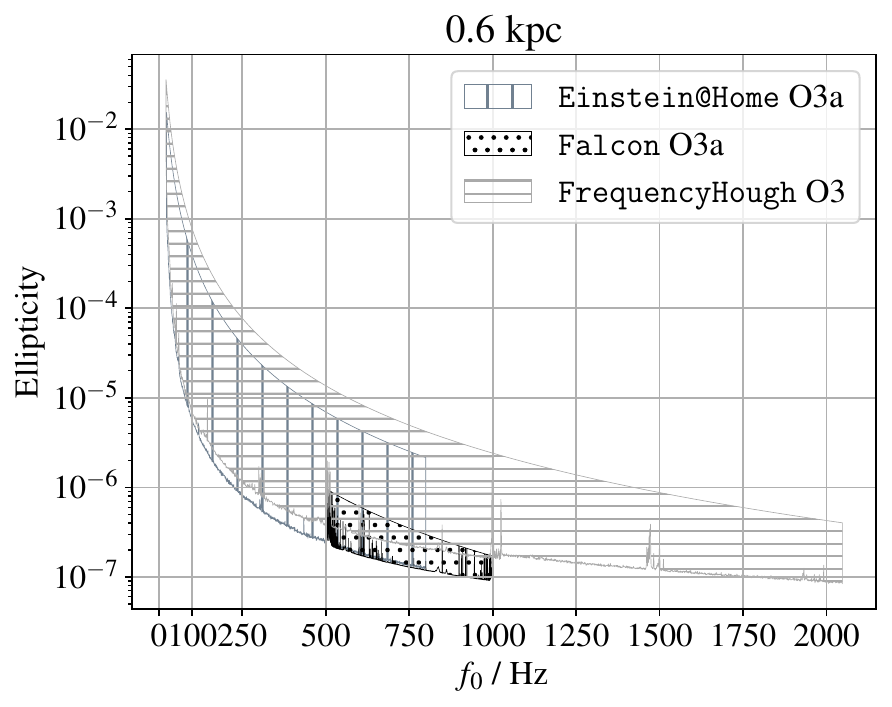}
    \includegraphics[width=\columnwidth]{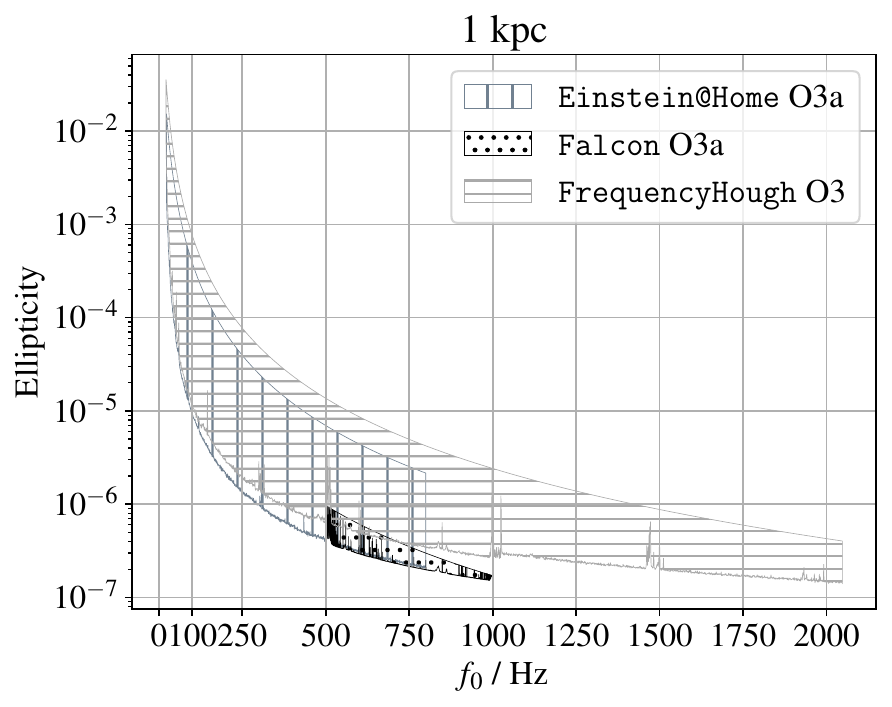}
    \caption{
        Ellipticity exclusion regions for the three searches considered in this work.
        The upper bound corresponds to $\ellipsd$ and is the same for different distances,
        whereas the lower bound corresponds to $\ellipul$ and increases with increasing distance.
    }
    \label{fig:O3_ul}
\end{figure}

We consider ellipticity exclusion regions from three all-sky searches:
the \texttt{FrequencyHough} results from the full O3 LVK search~\citep{KAGRA:2022dwb},
the \texttt{Einstein@Home} search in the first half of O3 data (O3a)~\citep{Steltner:2023cfk},
and the \texttt{Falcon} search in O3a~\citep{Dergachev:2022lnt}.
Frequency and spindown ranges are listed in Table~\ref{table:searches}.
Ellipticity upper limits are shown in Fig.~\ref{fig:O3_ul} for fiducial distances of
$\SI{600}{\parsec}$ and $\SI{1}{\kilo\parsec}$. 

The maximum expected ellipticity sustained by a typical NS is about
$\pellip{-6}{-5}$~\cite{Morales:2022wxs,Gittins:2021zpv}.
Ellipticities of $\ellip \approx 10^{-5}$ are excluded between
$\SI{100}{\hertz}$ to about $\SI{500}{\hertz}$.
Ellipticities of $\ellip\approx 10^{-6}$ are excluded from $\SI{250}{\hertz}$
to about $\SI{1500}{\hertz}$.
Ellipticities of $\ellip \lesssim 10^{-7}$ cannot be excluded for the distances
considered here.

The significance of these results depends on the population of sources under analysis,
which is typically modelled akin to the Galactic pulsar population~\cite{Reed:2021scb,Pagliaro:2023bvi} 
As discussed in Sec.~\ref{sec:introduction}, these searches are well suited 
to place constraints on the nearby population of \emph{unknown} sources
whose properties may differ significantly from those of pulsars.

\begin{table}
    \begin{tabular}{lrrr}
    \toprule
    Search & $\sdown{0}{}\,/ \unit{\hertz}$ & $-\sdown{1}{max}\,/\unit{\hertz^2}$ & Ref. \\
    \midrule
    \texttt{FrequencyHough} O3 & [10, 2048] & $1 \times 10^{-8}$ & \cite{KAGRA:2022dwb}\\
    \texttt{Einstein@Home} O3a & [20, 800] & $2.6\times 10^{-9}$ & \cite{Steltner:2023cfk}\\
    \texttt{Falcon} O3a & [500, 1000] &$5 \times 10^{-11}$ & \cite{Dergachev:2022lnt}\\
    \bottomrule
    \end{tabular}
    \caption{
        Parameter-space regions covered by the searches considered in this work.
    }
    \label{table:searches}
\end{table}

\section{Modelling young sub-kiloparsec gravitars}
\label{sec:gravitar_population}
The galactic NS population, in the context of CW searches,
has been extensively modelled under different assumptions
on the dominant emission mechanism~\cite{Palomba:2005na,
Knispel:2008ue,Wade:2012qc, Cieslar:2021viw, Reed:2021scb,
Pagliaro:2023bvi}. The latest detectability 
prospects~\cite{Pagliaro:2023bvi}, partly built upon the
pulsar population, suggest a reduced number of \emph{galactic}
sources ($\lesssim 1$) to be detectable by current detectors;
these results improve by an order of magnitude for 3G detectors.

In this section we construct a model for the sub-kiloparsec population
of unknown sources. To benefit from the astrophysical priors
proposed in~\cite{Popov:2003iq, Popov:2003hq,
2010AN....331..349H, 2014AN....335..935S},
we will assume a population of NSs younger than \SI{4}{\mega\year}
whose main spindown mechanism is the emission of CW,
i.~e.~\emph{gravitars}~\cite{Palomba:2005na}.
These sources emit at the spindown-limit amplitude~\cite{LIGOScientific:2007leh};
thus, as discussed in~\cite{Knispel:2008ue,Wade:2012qc}, the lack of a detection in future
observing runs will place constraints on the
sub-kiloparsec population of young NSs. Incidentally, these results will also constrain
the actual population of sub-kiloparsec young gravitars, whose plausible existence is discussed
in~\cite{Palomba:2005na,Knispel:2008ue}. Note also that older objects,
such as millisecond pulsars~\cite{Lorimer:2008se}, are also included in this work;  in such case,
``age'' refers to the spanned time since the considered object started to behave like a gravitar.

We assume CW emission from gravitars due to an equatorial
ellipticity $\ellip$. The frequency of the CW emission at present time
is thus derived from the general torque equation using a braking index of
$n=5$~\cite{Knispel:2008ue}:
\begin{equation}
    f(t ; \tage \fB, \taug) =
    \fB \left( 1 + \frac{t + \tage}{\taug} \right)^{-1/4} \,.
    \label{eq:f_gravitar}
\end{equation}
Here, $\tage$ corresponds to the age of the gravitar and $\fB$ to the
CW frequency at which the gravitar was born.
$t \in [0, \Tobs]$ is to be identified with the time since the beginning of an observing run.
The spindown timescale of the gravitar
\begin{equation}
    \taug = \frac{\KAbeta}{\fB^4 \ellip^2} \,
    \label{eq:taug}
\end{equation}
is consistent with the definition given in~\cite{Knispel:2008ue}, where 
\begin{equation}
    \KAbeta = \KAbetaLong = \KAbetaFiducial \,.
\end{equation}

Since the typical duration of an observing run is
\mbox{$\Tobs \lesssim \numrange[range-phrase=-]{0.5}{2}\, \unit{\year}$}
(thus $t \lesssim \tage$),
most blind searches Taylor-expand Eq.~\eqref{eq:f_gravitar} into
the so-called ``IT$1$ model''~\cite{Dergachev:2020upb, Covas:2022rfg}
\begin{align}
    f(t)  = & f_0 + f_1 t  + \mathcal{O}(t^2)\label{eq:IT1}\,, \\
    &f_0  = \fB  \left(1 + \frac{\tage}{\taug}\right)^{-1/4}\label{eq:f0}\,,\\
    &f_1  = -\frac{1}{4}\frac{f_0}{\tage + \taug}\label{eq:f1}\,.
\end{align}
As discussed in~\cite{Krishnan:2004sv}, the validity of this model depends on
the age of the source: the older the source, the slower the spindown rate,
which limits the amount of relevant spindown terms for a given observing time.
This will place a limit on the minimum age of the population considered so that
Eq.~\eqref{eq:IT1} remains valid, to be discussed in Sec.~\ref{subsec:tage}.
The study of other frequency evolution models, such as higher order terms or
power-law models~\cite{Oliver:2019ksl}, are left for future work.

Throughout the following subsections, we discuss plausible prior
distributions for $\ellip$, $\tage$, and $\fB$.
The resulting population is presented in Sec.~\ref{subsec:population}.

\subsection{Ellipticity distribution}
\label{subsec:ellip}

The typical ellipticity of a NS is highly uncertain
(see~\cite{Lasky:2015uia,Glampedakis:2017nqy} and references therein).
The latest upper bound on the ellipticity a standard NS could sustain is
\mbox{$\ellip \lesssim 10^{-5}$}~\cite{Gittins:2021zpv, Morales:2022wxs},
although higher values are possible for exotic equations of 
state~\cite{Owen:2005fn}. A minimum ellipticity of 
\mbox{$\ellip\approx 10^{-9}$} is suggested for
the observed population of millisecond pulsars~\cite{Woan:2018tey}.
We take $\logellip$ to be uniformly distributed along $[-9, -5]$
to cover the upper end of plausible values for the nearby
gravitar population. Specifically, we will quote results for four
different ellipticity ranges: $[-9, -8]$, $[-8, -7]$, $[-7, -6]$, $[-6, -5]$.

\subsection{Age distribution and the linear spindown model}
\label{subsec:tage}

We chose $\tage$ to be uniformly distributed
between $\SI{5}{\kilo\year}$ and $\SI{4}{\mega\year}$.
The upper end, as discussed earlier in this section, owes to the
astrophysical priors suggested
in~\cite{Popov:2003iq, Popov:2003hq, 2014AN....335..935S}.
The lower end is chosen so that the model in Eq.~\eqref{eq:IT1}
is valid for all the ellipticity values considered in Sec.~\ref{subsec:ellip}.
We consider Eq.~\eqref{eq:IT1} to be valid if the contribution of the 
quadratic term to the phase evolution of the signal along an observing run
is lower than a quarter of a cycle~\cite{Jaranowski:1998qm, Krishnan:2004sv}.

\subsection{Birth frequency distribution}
\label{subsec:fB}

The distribution of $\fB$ for the Galactic NS population is uncertain,
and different estimates can be constructed using the observed NS population
or numerical simulations of core-collapse supernova. 
We refer the reader to~\cite{Pagliaro:2023bvi} and references
therein for a review on the latest proposals.
To reflect our ignorance in $\fB$ we chose two broad uniform 
distributions along the sensitive band of ground-based interferometric
detectors.

The first distribution, labeled as ``low birth-frequency'',
spans from $\SI{50}{\hertz}$ to \SI{1}{\kilo\hertz}.
The lower limit is consistent with the physical arguments
given in~\cite{Palomba:2005na} supporting the existence of gravitars.
The upper end corresponds to a rotational frequency of~\SI{500}{\hertz}
and is broadly consistent with the upper end of most astrophysical distributions
on $\fB$ discussed in~\cite{Palomba:2005na, Knispel:2008ue, Pagliaro:2023bvi}.

The second distribution, labeled as ``high birth-frequency'', spans
from $\SI{1}{\kilo\hertz}$ to $\SI{2}{\kilo\hertz}$ and covers the
highest frequency values analyzed by a CW search to date. The maximum
value corresponds to a rotational frequency of $\SI{1}{\kilo\hertz}$ and
is below the Keplerian break-up frequency of a NS~\cite{Haskell:2018nlh}.
This frequency range is broadly consistent with the expected emission frequency
of NSs in accreting systems, which are promising CW emitters~\cite{Gittins:2018cdw}.

\subsection{Simulating a gravitar population}
\label{subsec:population}

\begin{figure}
    \includegraphics[width=\columnwidth]{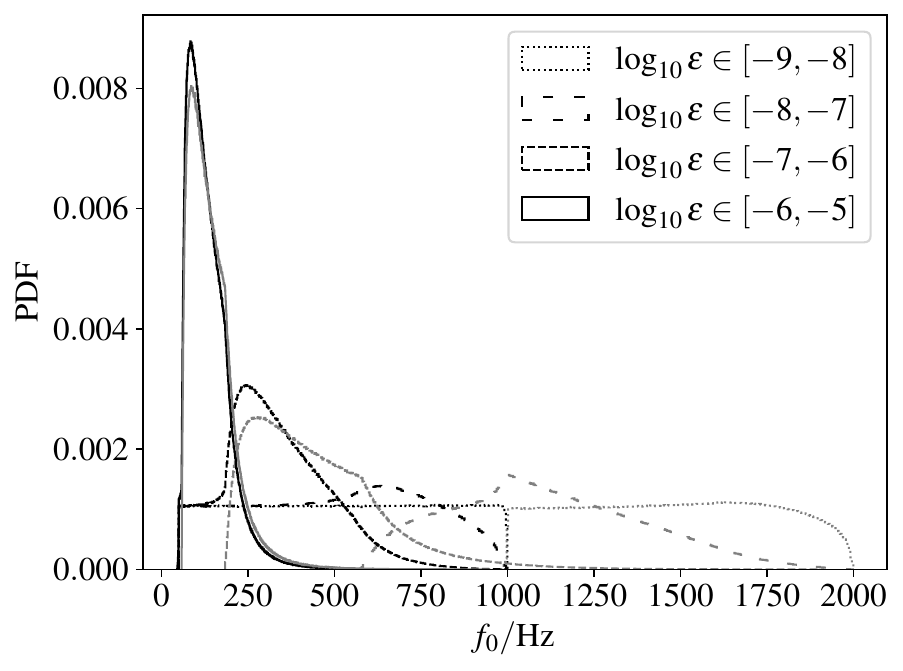}
    \includegraphics[width=\columnwidth]{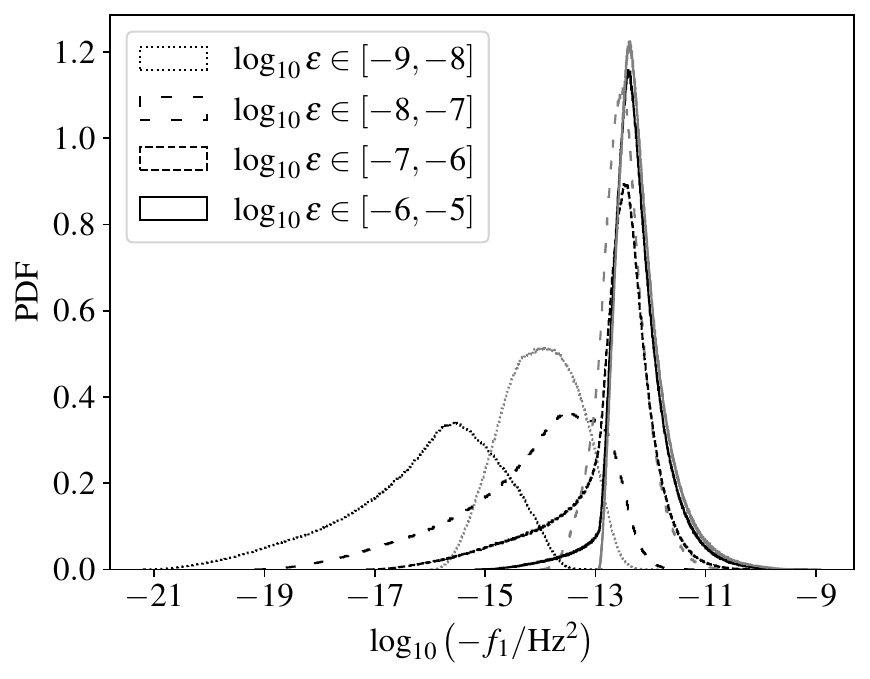}
    \caption{
        Distribution of $f_0$ and $f_1$ for a gravitar population
        at present time as described in Sec.~\ref{sec:gravitar_population}.
        Black histograms correspond to the low birth-frequency distribution; gray
        histograms correspond to the high birth-frequency distribution (see Sec.~\ref{subsec:fB}).
        Each range of frequencies and ellipticities contains $10^{7}$ samples.
        Each sample corresponds to sampling $\ellip$, $\tage$ and $\fB$ from the specified
        priors and computing $f_0$ and $f_1$ using Eqs.~\eqref{eq:f0}~and~\eqref{eq:f1}.
        }
        \label{fig:f0}
\end{figure}

\begin{figure}
    \includegraphics[width=\columnwidth]{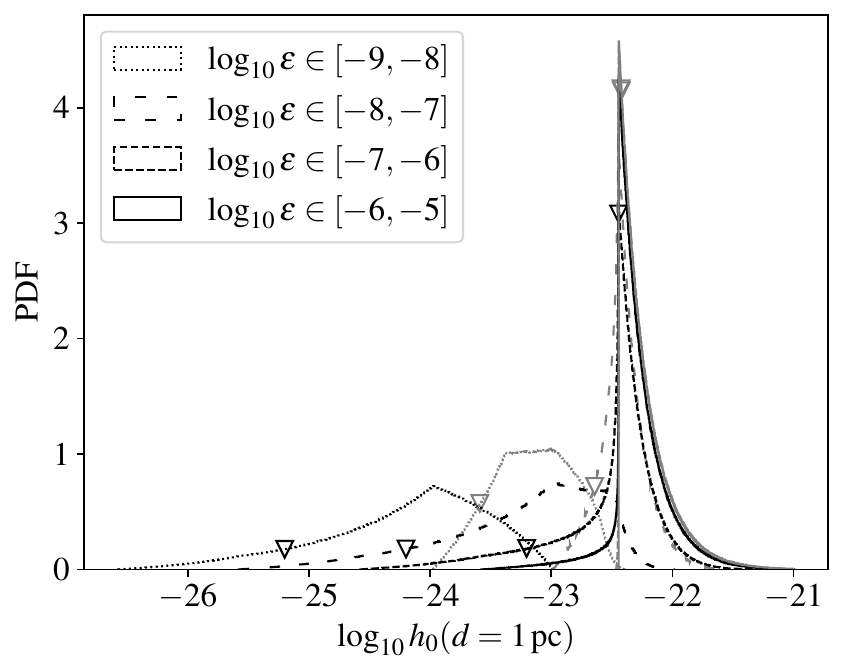}
    \caption{
        Distribution of $h_0$ for a gravitar population at \SI{1}{\parsec}
        at present time as described in Sec.~\ref{sec:gravitar_population}.
        Black histograms correspond to the low birth-frequency distribution; gray
        histograms correspond to the high birth-frequency distribution (see Sec.~\ref{subsec:fB}).
        Each range of frequencies and ellipticities contains $10^{7}$ samples.
        Each sample corresponds to sampling $\ellip$, $\tage$ and $\fB$ from the specified
        priors and computing $h_0$ using Eq.~\eqref{eq:h0} for $d=\SI{1}{\parsec}$ and
        $I = \SI{1e38}{\kilo\gram \meter^2}$.
        Triangular markers denote the $10\%$ percentile of each histogram.
    }
    \label{fig:h0}
\end{figure}

For a given gravitar population (i.e. birth-frequency and ellipticity distributions),
we sample the parameters of $N_{\star} = 10^{7}$ gravitars
\begin{equation}
    (\ellip^{(j)}, \tage^{(j)}, \fB^{(j)}) \sim p(\ellip, \tage, \fB) ,\, j = 1, \dots, N_{\star},
\end{equation}
where $p(\ellip, \tage, \fB)$ represents the distribution of gravitar parameters for the
selected population.
We then use Eq.~\eqref{eq:f0} to compute the CW emission frequency of these gravitars
at present time
\begin{equation}
    f_0^{(j)} = f_0(\ellip^{(j)}, \tage^{(j)}, \fB^{(j)})\,,
\end{equation}
and Eq.~\eqref{eq:f1} to compute their spindown
\begin{equation}
    f_1^{(j)} = f_1(\ellip^{(j)}, \tage^{(j)}, \fB^{(j)})\,.
\end{equation}
The CW amplitude of a gravitar at a distance $d$ is given by Eq.~\eqref{eq:h0}
\begin{equation}
    h_0^{(j)}(d) = h_0(f_0^{(j)}, \ellip^{(j)}, d) \,.
\end{equation}
The resulting  $f_0$, $f_1$, and $h_0$ distributions
are shown in Fig.~\ref{fig:f0} and Fig.~\ref{fig:h0}.
We summarize their main features to conclude this discussion.

We start by explaining the local maxima in the $f_0$ histograms in Fig.~\ref{fig:f0},
which were also discussed in~\cite{Knispel:2008ue} in a slightly different context.
To do so, let us introduce the following variables:
\begin{align}
    \ptaug &= \frac{\taug}{\tage}\,,\\
    \pt &= \frac{t}{\tage}\,.
\end{align}
Note that \mbox{$\pt \lesssim \mathcal{O}(10^{-4})$} since
$\max_{t} t = \Tobs \leq (1-2) \unit{\year}$
and $\tage \gtrsim \SI{5}{\kilo\year}$. 
On the other hand, $\ptaug$ may or may not be small, depending
on the ellipticity and birth frequency of the considered gravitar population.
Using these variables, we expand Eq.~\eqref{eq:f_gravitar} to zeroth
order in $\pt$
\begin{align}
    f(t ; \tage, \fB, \ellip) 
    & \simeq  \fauld(\tage, \ellip)  \left(1 + \ptaug\right)^{-1/4}
    \label{eq:f_expanded}
\end{align}
where
\begin{equation}
    \fauld(\tage, \ellip) = \left( \frac{\KAbeta}{\tage \ellip^2} \right)^{1/4} \,
    \label{eq:fauld}
\end{equation}
and the expansion error is at most
\begin{equation}
    \begin{split}
    \frac{1}{4}\fauld(\tage, \ellip)  \left(1 + \ptaug\right)^{-5/4} \pt =& \\
    \SI{2.5e-2}{\hertz}
    \left( \frac{\fauld(\tage, \ellip)}{\SI{1}{\kilo\hertz}} \right)
    \left(1 + \ptaug\right)^{-5/4}  
    \left(\frac{\pt}{10^{-4}}\right) &\,.
    \end{split}
\end{equation}
The dependency of Eq.~\eqref{eq:f_expanded} on $\fB$ is entirely contained
in $\ptaug$. In the limit $\ptaug \rightarrow 0$ (which corresponds to $\tage \gg \taug$),
the frequency of a gravitar is entirely determined by $\tage$ and $\ellip$
(and \emph{not} $\fB$) within a fraction of a hertz. 
Thus, $\taug$ is the typical timescale accounting for the dependency of $f(t)$
on the gravitar's birth frequency $\fB$. Conversely, for a given ellipticity $\ellip$
and age $\tage$, $\fauld(\tage, \ellip)$ marks the frequency beyond which
no gravitar will be found, as any higher birth frequency would have spun down below
$\fauld$ after $\tage$. The maxima in the $f_0$ histogram thus correspond to
$\fauld(\tage, \ellip)$ for $\tage \approx \SI{4}{\mega \year}$ and maximum ellipticity,
as for those gravitars $\tage \gg \taug$.

For the selected range of ages, gravitars with \mbox{$\logellip \gtrsim -8$} tend to fall into
the \mbox{$\tage \gg \taug$} regime, while gravitars with \mbox{$\logellip \lesssim -8$} 
tend to be distributed more consistently with the birth frequency distribution. Specifically, the
\mbox{$\logellip \in [-6, -5]$} population is contained below
\mbox{$f_0 \lesssim \SI{500}{\hertz}$};
similarly the $\logellip \in [-7, -6]$ population is practically contained below
$f_0 \lesssim \SI{1250}{\hertz}$. We note that high-ellipticity sources tend to be less affected
by the chosen birth-frequency distribution and are found along the most sensitive band of the detectors.

The spindown values of these populations are
\mbox{$\left| f_1 \right| \lesssim \SI{1e-9}{\hertz^2}$}, {
which is well within the spindown range covered by the
searches discussed in Sec.~\ref{sec:ellipticity_from_uls} 
(see Table~\ref{table:searches}).

Crossing the results in Fig.~\ref{fig:O3_ul} and Fig.~\ref{fig:f0}
we can draw the following conclusion: ellipticity exclusion upper
limits for $\ellip \approx 10^{-5}$ are valid up to \mbox{$f_0 \approx \SI{500}{\hertz}$},
as \mbox{$\ellipsd(\SI{500}{\hertz})\approx 10^{-5}$}; 
As a result, the searches here discussed cover the totality of the $\ellip \approx 10^{-5}$
gravitar population and exclude their presence within $\SI{1}{\kilo\parsec}$ down to \SI{100}{\hertz}.
This is because the gravitar population's spindown range is covered by the searches's,
and is valid for any other ellipticity $\ellip < 10^{-5}$.
Similarly, for $\ellip \approx 10^{-6}$, the exclusion is valid down to about \SI{250}{\hertz}. 
The upper limits are not sensitive enough to place constraints at $\SI{1}{\kilo\parsec}$
for lower ellipticities.

\section{Sensitivity of blind CW searches to sub-kiloparsec sources}
\label{sec:sensitivity}
In this section we introduce the quantities that will be relevant to evaluate
the detectability of the sub-kiloparsec populations of gravitars.
Extended discussion of these results will be presented
in Sec.~\ref{sec:discussion}.

We quantify the sensitivity of a CW search using the 
sensitivity depth~\cite{Behnke:2014tmat, Dreissigacker:2018afk}
\begin{equation}
    \depth = \frac{\sqrtSn}{h_0}\,,
    \label{eq:depth}
\end{equation}
Here, $h_0$ represents the average amplitude detectable by a search
at a certain confidence level
(assuming a specific signal population) and $\sqrtSn$ represents the single-sided amplitude
spectral density (ASD) of the noise.

\begin{figure}
    \includegraphics[width=\columnwidth]{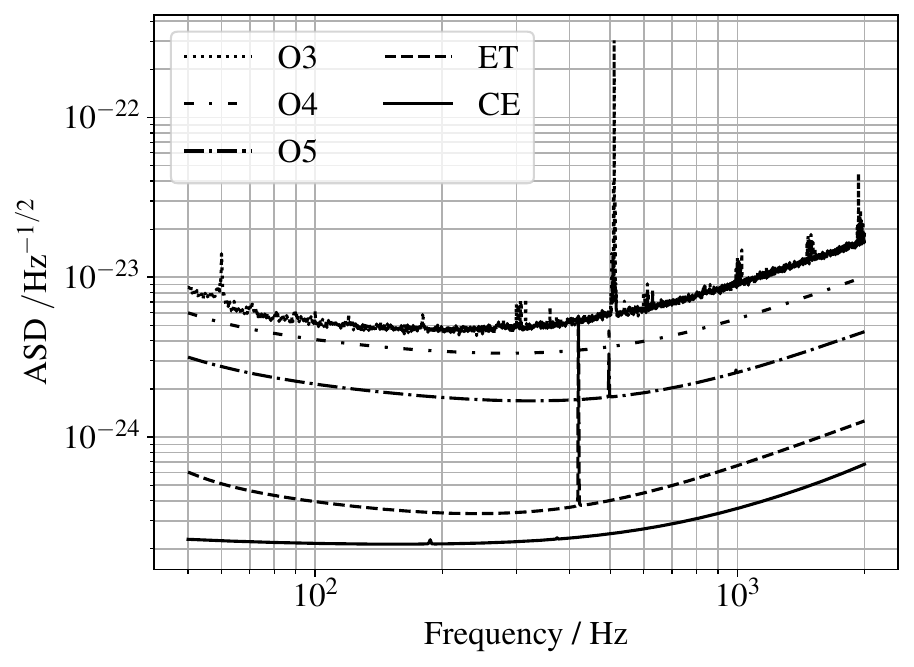}
    \caption{
        Detector noise curves for the [50, 2000] \unit{\hertz} frequency band of the
        Advanced LIGO and thrid-generation ground-based detectors. O3 corresponds to
        a harmonic average of representative
        ASDs~\cite{O3aH1ASD,O3aL1ASD,O3bH1ASD,O3bL1ASD}.
        O4~\cite{O4ASD}, O5~\cite{O5ASD}, ET~\cite{ETASD,Hild:2009ns}, 
        and CE~\cite{CEASD,Evans:2021gyd} correspond to projected sensitivities.
        We assume a single CE detector with a \SI{40}{\kilo\meter} length.
    }
    \label{fig:asd}
\end{figure}

We will evaluate the sensitivity of blind CW searches for different detector configurations
whose ASDs are shown in Fig.~\ref{fig:asd}.
In a similar fashion to \cite{Moragues:2022aaf, Pagliaro:2023bvi}, 
this requires the introduction of a prefactor accounting for the number of interferometric
detectors $N_{\mathrm{IFO}}$ and their arm angle $\zeta$.
For the Advanced LIGO detectors~\cite{AdvancedLIGO} 
\mbox{($N_{\mathrm{IFO}}=2, \sin\zeta=1$)}
we consider the third (O3), fourth (O4), and fifth (O5) observing runs.
The ASD for O3 is computed as the harmonic average of representative per-detector
ASDs~\cite{O3aH1ASD, O3aL1ASD, O3bH1ASD, O3bL1ASD}.
For O4 and O5, we use the ASDs provided in~\cite{O4ASD} and~\cite{O5ASD}, respectively.
For 3G detectors, we take ET~\cite{Maggiore:2019uih} to be composed by three interferometers
with \SI{60}{\degree} arms \mbox{$(N_{\mathrm{IFO}}=3, \sin\zeta=\sqrt{3}/2)$}
using the ASD provided in~\cite{Hild:2009ns, ETASD};
we assume CE~\cite{Reitze:2019iox} to be composed of a single detector
\mbox{$(N_{\mathrm{IFO}}=1, \sin\zeta=1)$} with the ASD given in~\cite{CEASD,Evans:2021gyd}.

Estimating the sensitivity of a search, albeit possible~\cite{Dreissigacker:2018afk},
is a complicated endeavour due to the abundance of configuration choices with complicated behavior.
We take a similar approach to~\cite{Pagliaro:2023bvi} and instead take the sensitivity depth of
the latest CW searches~\cite{KAGRA:2022dwb, Steltner:2023cfk, Dergachev:2022lnt}
as a proxy for a representative all-sky CW search:\footnote{
    Typical searches quote sensitivity depth at a 90\% or 95\% confidence level~\cite{Tenorio:2021wmz}. 
    The variation caused by this is well within the order of magnitude of the results derived here.
}
\begin{equation}
    \depth(f_0) = \sin{\zeta} \sqrt{ \frac{N_{\mathrm{IFO}}}{2} } \times
    \begin{cases}
        55 / \sqrt{\unit{\hertz}} & f_0  <  \SI{1}{\kilo\hertz} \\
        30 / \sqrt{\unit{\hertz}} & f_0  >  \SI{1}{\kilo\hertz}
    \end{cases}\,.
    \label{eq:depth_searches}
\end{equation}
This choice reflects the current trend in all-sky searches,
where sensitivity depth is lower at higher frequencies due the rapidly
increasing trials factor. With this, we can compute the detectable $h_0$
for different observing runs and detector configurations along a frequency band.
The resulting $h_0(f)$ is to be compared with the results in Fig.~\ref{fig:h0}
at the appropriate distance.

\subsection{Detectablity distance}
\label{subsec:distance}

We are interested in computing the \emph{detectability distance} of the gravitar
populations discussed in Sec.~\ref{sec:gravitar_population}. This corresponds to the
distance at which a certain fraction of a gravitar population can be detected by
search pipeline using a specific detector configuration. 

At a given distance $d$, and given a detector configuration
\mbox{$(\sqrtSn, N_{\mathrm{IFO}}, \sin\zeta)$},
the detectable fraction of a gravitar population $\pdet(d)$
is computed using Monte Carlo integration as~\cite{Searle:2008jv}
\begin{equation}
    \pdet(d) = \frac{1}{N_{\star}} \sum_{j=1}^{N_{\star}} 
    \left\llbracket
        h_0^{(j)}(d) > \frac{\sqrtSn(f_0^{(j)})}{\depth(f_0^{(j)})}
        \right\rrbracket \,,
\end{equation}
where $\depth(f_0)$ is given in Eq.~\eqref{eq:depth_searches} and
the Iverson bracket $\llbracket \cdot \rrbracket$ evaluates to 1 (0) whenever
the expression inside them is true (false). The detectability distances for different
detected fractions and gravitar populations using different detector configurations
are shown in Fig.~\ref{fig:distance_plot_low} and Fig.~\ref{fig:distance_plot_high}.
In general, the detectability distance increases by a factor
$1.2$ to $2$ between O3 and O4,
$2$ to $3$ between O3 and O5,
and $10$ to $20$ between O3 and 3G detectors.

\begin{figure*}
    \includegraphics[width=\mywidth]{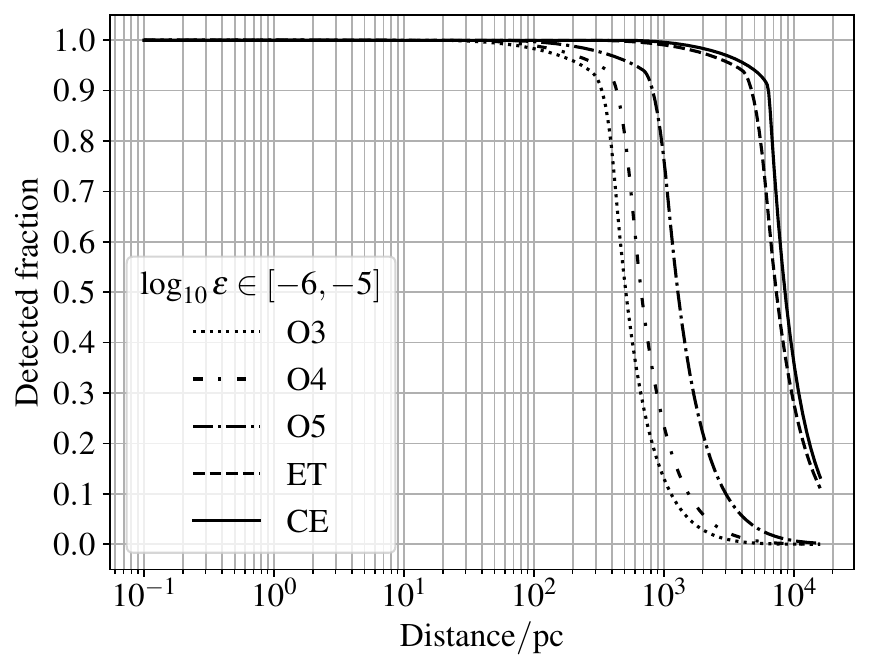}
    \includegraphics[width=\mywidth]{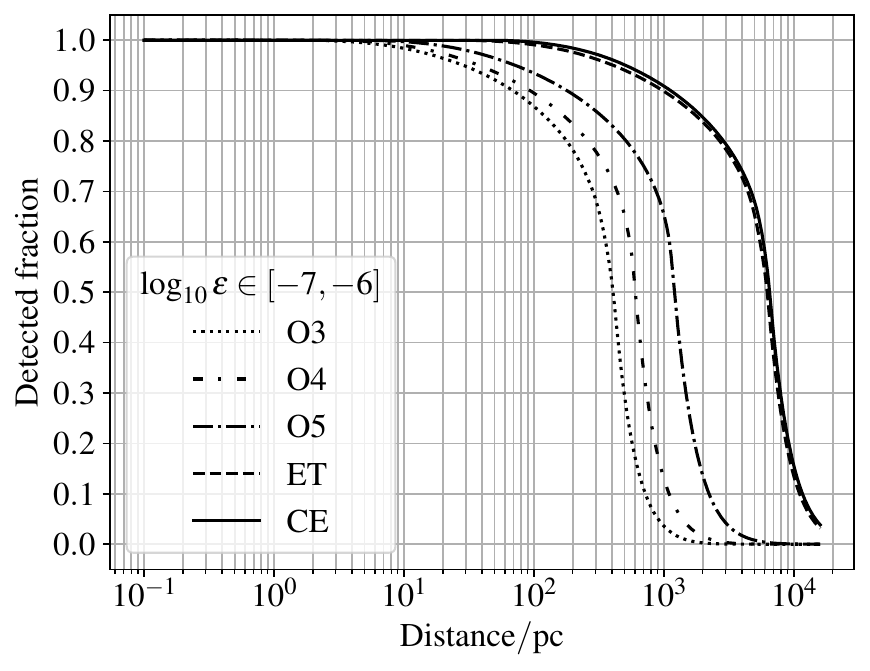}
    \includegraphics[width=\mywidth]{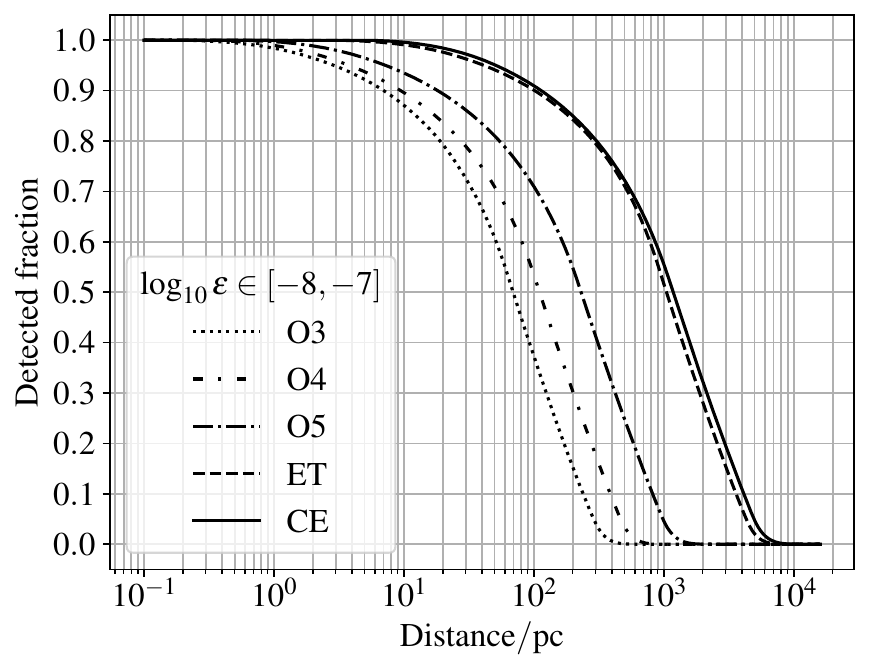}
    \includegraphics[width=\mywidth]{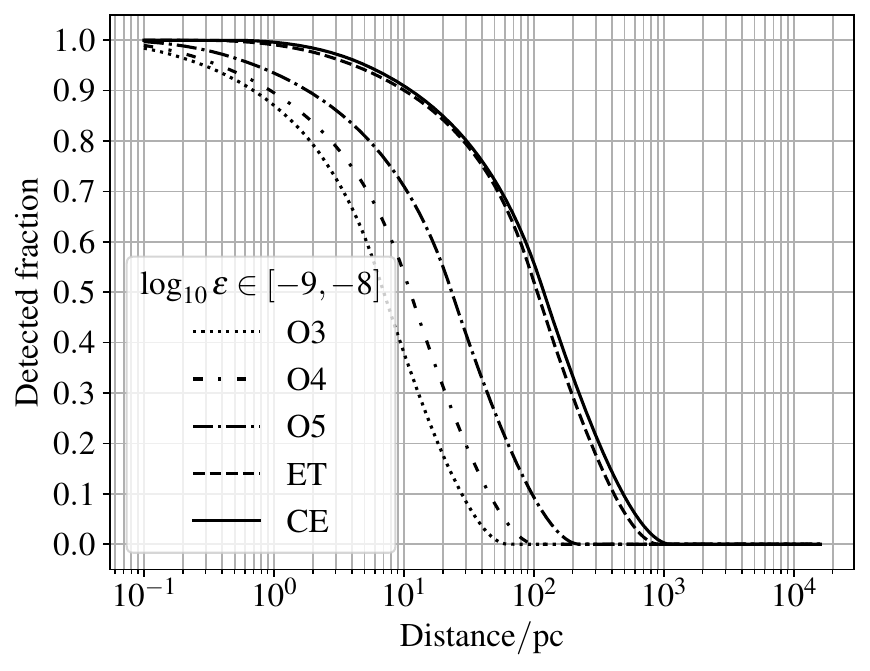}
    \caption{
        Detectability distance for detected fractions assuming a low birth-frequency
        gravitar population $\fB \in [50, 1000] \unit{\hertz}$ for different ellipticities and
        detector configurations.
    }
    \label{fig:distance_plot_low}
\end{figure*}

\begin{figure*}
    \includegraphics[width=\mywidth]{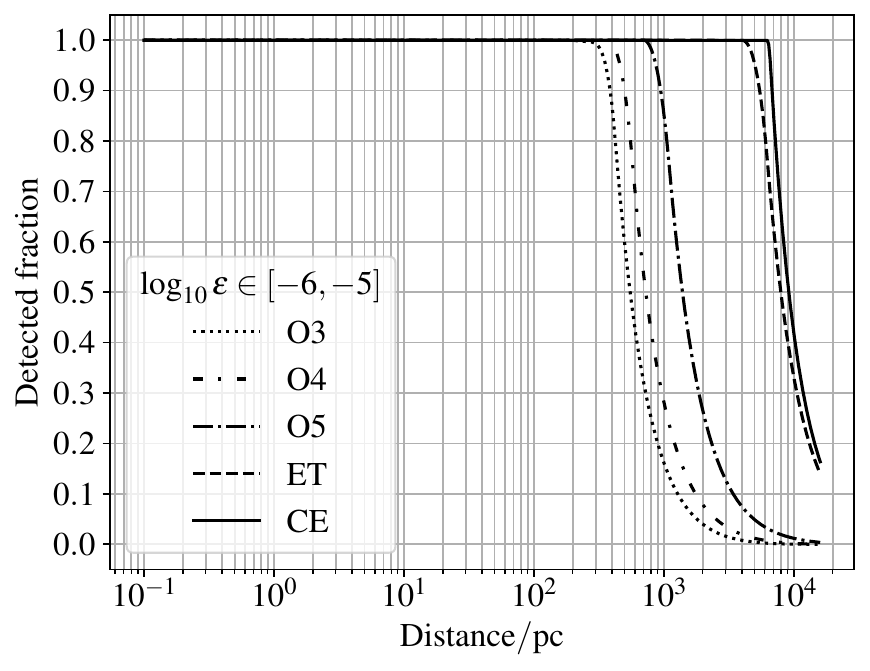}
    \includegraphics[width=\mywidth]{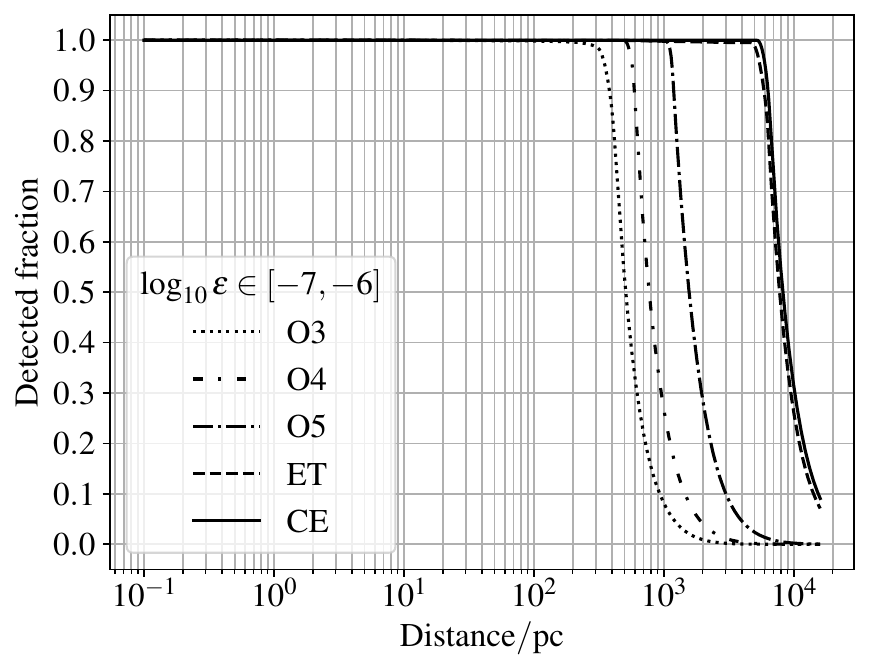}
    \includegraphics[width=\mywidth]{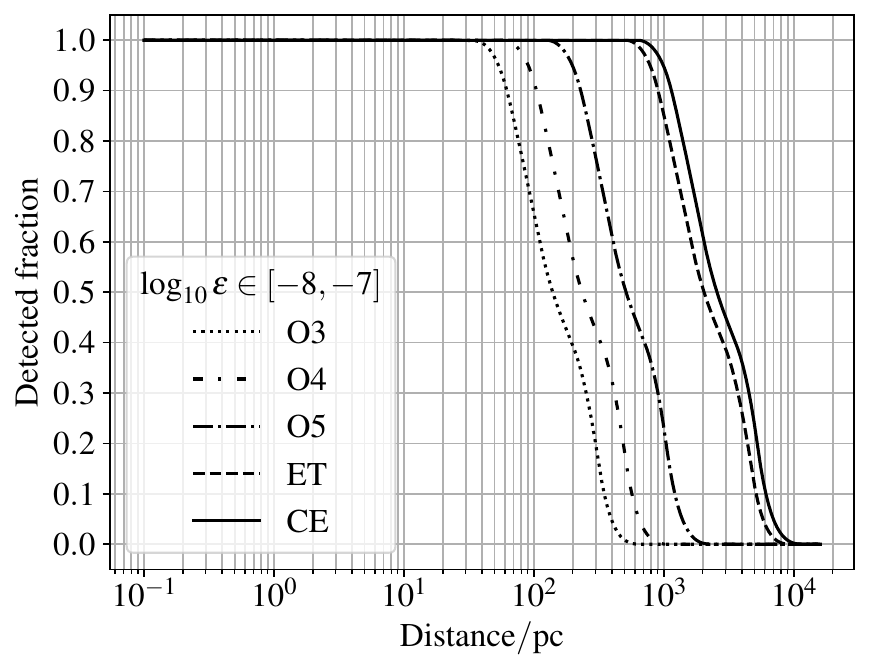}
    \includegraphics[width=\mywidth]{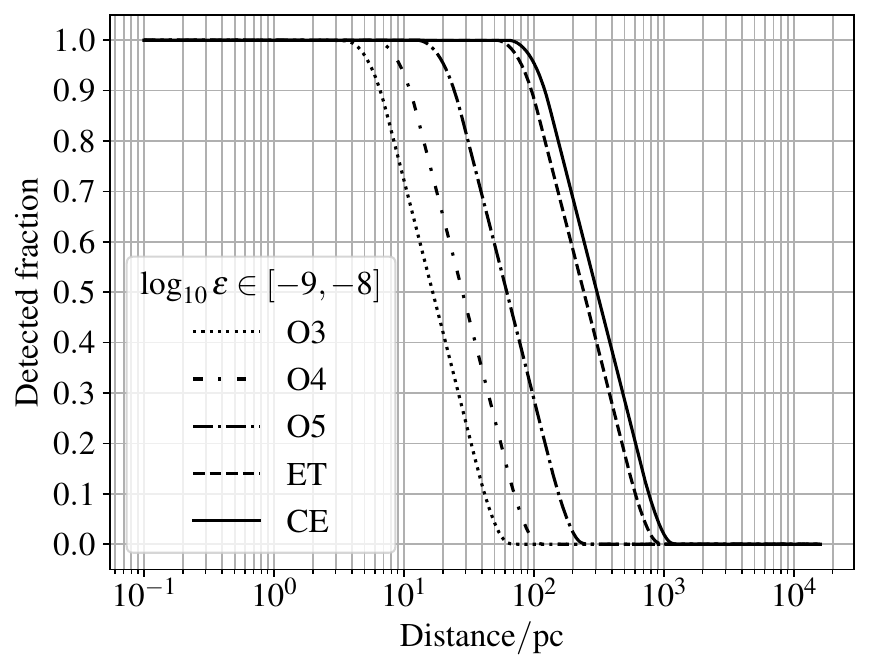}
    \caption{
        Equivalent to Fig.~\ref{fig:distance_plot_low}, but with $\fB \in [1000, 2000] \unit{\hertz}$.
    }
    \label{fig:distance_plot_high}
\end{figure*}

The population of sub-kiloparsec NSs younger than \SI{4}{\mega\year}
is expected to contain on the order of 100 objects~\cite{Popov:2003iq,Popov:2003hq}.
Out of those, an unknown (and possibly small) fraction
may present ellipticities within the ranges considered in this work.
To take this low number of sources into account, we collect in Table~\ref{table:distance_90}
the detectability distances for different detectors and gravitar populations so that
90\% of the population is detected $d^{90\%}$.
This quantity serves as a proxy for the astrophysical reach of a blind search using
a specific detector configuration.

\subsection{Constraining the number of sub-kiloparsec gravitars}
\label{subsec:number}

If no CW signals are detected, an upper bound on the expected
number of sub-kiloparsec gravitars can be derived as follows.
Let us assume a uniform spatial density of gravitars $\rhog$
across a nearby volume $\Vt$.
The expected number of gravitars $\Ng$ within $\Vt$ is given by 
\begin{equation}
    \Ng = \rhog \Vt\,.
\end{equation}
The lack of observed sources within a distance $d$
probes a fraction $V(d)$ of the volume of interest $\Vt$.
$V(d)$ corresponds to the \emph{intersection} of a sphere
of radius $d$ centered at the detector with $\Vt$.
The density of gravitars in $\Vt$ is thus bounded by 
\begin{equation}
    \rhog \lesssim \frac{1}{V(d)} \,.
\end{equation}
Consequently, the number of gravitars within $V$ is bounded by
\begin{equation}
    \Ng \lesssim \frac{\Vt}{V(d)} \,.
    \label{eq:Ng_bound}
\end{equation}

We first take $\Vt$ to be the galactic stellar disk (GD) up to \SI{0.6}{\kilo\parsec}
and \SI{1}{\kilo \parsec} away from the detector. For a given distance $d$,
the surveyed volume $V(d)$ corresponds to the intersection
of a sphere center at the detector with the GD.
This can be computed as~\cite{Wade:2012qc}
\begin{equation}
    V(d) = \begin{cases}
        \frac{4}{3} \pi d^3 & d < H \\
        \frac{2}{3} \pi H (3 d^2 - H^2) & d > H
    \end{cases}\,,
\end{equation}
where $H = \SI{75}{\parsec}$ is half the height of the GD.
The resulting $\Ng$, for both maximum distances, is represented 
in Fig.~\ref{fig:number_gravitars} using a stripe. For reference,
we include in Fig.~\ref{fig:number_gravitars} the expected number
of nearby NSs using the lower-end estimate of
\mbox{$10^{-4}\,\mathrm{NS} \, \mathrm{pc}^{-3}$}
provided in~\cite{2010A&A...510A..23S}. This estimate includes NSs
beyond the population considered in this work.

\begin{table}
    \begin{tabular}{rrrrr}
        \multicolumn{5}{@{}c}{$\fB \in [50, 1000] \unit{\hertz}$}\\ 
        \toprule
        & \multicolumn{4}{@{}c}{$\log_{10}{\ellip}$}\\
            \cmidrule(r){2-5}
        &$ [-9, -8]$ & $[-8, -7]$ & $[-7, -6]$ & $[-6, -5]$\\
        \midrule
        O3 & 0.7 (0.1) & 7 (1) & 70 (10) & 330 (10) \\
        O4 & 1.0 (0.1) & 10 (1) & 90 (10) & 430 (10)\\
        O5 & 1.8 (0.3) & 18 (3) & 180 (30)  & 810 (20)    \\
        ET & 10 (1) & 100 (10) & 1000 (100) & 4700 (100)\\
        CE & 11 (1) & 110 (10) & 1100 (100) & 6420 (80)\\
        \bottomrule
    \end{tabular}
    \begin{tabular}{rrrrr}
        \multicolumn{5}{@{}c}{$\fB \in [1000, 2000] \unit{\hertz}$}\\ 
        \toprule
        & \multicolumn{4}{@{}c}{$\log_{10}{\ellip}$}\\
            \cmidrule(r){2-5}
        &$ [-9, -8]$ & $[-8, -7]$ & $[-7, -6]$ & $[-6, -5]$\\
        \midrule
        O3 & 6.5 (0.2)   & 62 (1)     & 386 (5)      & 390 (10) \\
        O4 & 11.0 (0.4)  & 105 (3)   & 596 (1)         & 503 (6) \\
        O5 & 24.4 (0.5) & 228 (5)  & 1196 (1)         & 940 (11) \\
        ET & 96 (2)       & 900 (20)& 5980 (70) & 5430 (70) \\
        CE & 122 (3)        & 1140 (30)& 6270 (80) & 6740 (80) \\
        \bottomrule
    \end{tabular}
    \caption{
        Distance (in parsec) at which $90\%$ of the gravitar population
        can be detected for the specified detector configuration, ellipticity,
        and birth-frequency distribution.
        Figures in brackets correspond to absolute uncertainty, which we take to be
        half the separation between $89\%$ and $91\%$ distances.
    }
    \label{table:distance_90}
\end{table}

\begin{figure}
    \includegraphics[width=\columnwidth]{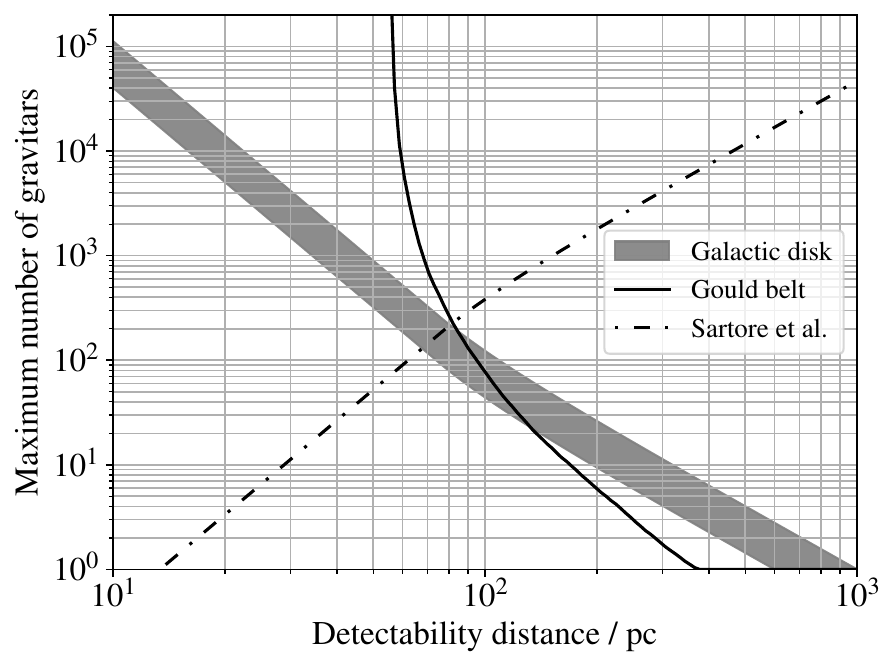}
    \caption{
        Estimated upper bounds on the number of sub-kiloparsec graviars for different 
        spatial distributions of the population. For a given detectability distance,
        the vertical axis represents an upper bound on the expected number of gravitars
        from the density estimate computed using Eq.~\eqref{eq:Ng_bound}.
        The shaded region assumes a uniform distribution of gravitars in the galactic disk
        (GD) up \SI{0.6}{\kilo\parsec} and \SI{1}{\kilo\parsec} away from the Earth.
        The solid line assumes a uniform distribution of gravitars across the Gould belt (GB). 
        Additionally, the dashed-dotted line shows an estimate on the number 
        of nearby NSs assuming a density of \SI{1e-4}{\parsec^{-3}}~\cite{2010A&A...510A..23S}.
        }
    \label{fig:number_gravitars}
\end{figure}

We also take $V$ to be the Gould belt (GB), which is proposed to harbor
an overabundance of unobserved NSs~\cite{Popov:2003hq, Popov:2003iq}.
To do so, we follow~\cite{Palomba:2005na}
and model the GB as a hollow cylinder with an inner radius of $\SI{150}{\parsec}$,
an outer radius of $\SI{500}{\parsec}$, and a height of $\SI{60}{\parsec}$.
The cylinder is centered $\SI{100}{\parsec}$ away from the Sun
towards the antigalactic direction, and is tilted $\SI{18}{\degree}$
with respect to the galactic plane. 
In this case, $\Vt \approx \SI{4.3e7}{\parsec^{3}}$ is computed as 
the volume of a hollow cylinder. We compute $V(d)$ as the intersection of a
sphere of radius $d$ center at the detector with the GB using a Monte Carlo
integral with $10^5$ samples, which returns a negligible uncertainty.
The results are shown as a solid line in Fig.~\ref{fig:number_gravitars}.
This approach can be applied to any other galactic structures of interest,
such as the recently-proposed Radcliff wave~\cite{2020Natur.578..237A};
the discussion of other nearby structures potentially harbouring undetected NSs
is left for future work.

\section{Discussion}
\label{sec:discussion}
With the results presented in Sec.~\ref{sec:sensitivity} we estimate A) which gravitar
populations could be constrained by all-sky searches and B) how much these constraints
could improve if alternative search strategies were used.

\subsection{Constraints on sub-kiloparsec CW sources}

The detectability distances collected in Table~\ref{table:distance_90} are to be compared
with the estimated upper bounds on the number of sub-kiloparsec gravitars
[Eq.~\eqref{eq:Ng_bound}] shown in Fig.~\ref{fig:number_gravitars}. For example, if a search's
detectability distance is $d = \SI{10}{\parsec}$, the number of gravitars in the
\SI{0.6}{\kilo\parsec} (\SI{1}{\kilo\parsec}) GD is bounded by $4 \times 10^{4}$ ($10^{5}$).
At $d = \SI{100}{\parsec}$, this bound reaches $50$ ($100$) gravitars.
In order to bound the number of gravitars in the GB, the detectability distance needs to be
higher than \SI{50}{\parsec}, as otherwise the sphere centered at the detector does not
overlap with the GB. 

These bounds are astrophysically meaningful whenever they dive below the expected number of
NSs within the surveyed volume. As shown in Fig.~\ref{fig:number_gravitars},
the GD upper bounds dive below the estimated number of NSs~\cite{2010A&A...510A..23S}
at  $d \approx \SI{100}{\parsec}$, which roughly corresponds to an upper bound
of $\Ng \lesssim 100$. This would implie less than $0.25\%$ of the nearby population of NSs
is composed of these gravitars. For the GB, which is expected to contain about $100$
objects~\cite{Popov:2003iq,Popov:2003hq}, the upper bound becomes meaningful at
$d \approx \SI{100}{\parsec}$ as well. 

The population of $\pellip{-6}{-5}$ gravitars shows a similar behavior for both
birth-frequency distributions.
The results in O3 data constrain the population of GD gravitars to \mbox{$\Ng \lesssim 10$}.
O5 results will bring this constraint down to \mbox{$\Ng \lesssim 1$}.
For the case of the GB, the constraint will reach \mbox{$\Ng \lesssim 1$} already in O4.
High birth-frequency GD gravitars with  $\pellip{-7}{-6}$
are constrained to $\Ng \lesssim 10$ by O3 results.
Constraints at the level of \mbox{$\Ng \lesssim 1$}
will be reached for both the GD and GB populations after O4.
The low-birth frequency population will be constrained to \mbox{$\Ng \lesssim 100$} 
using O5 data. 
Constraints at the level of \mbox{$\Ng \lesssim 1$} will be put in place by 3G detectors.
For \mbox{$\pellip{-8}{-7}$}, the high birth-frequency population will be constrained 
down to \mbox{$\Ng \lesssim 100$} in O4; 3G detectors will bring this constraint down to 
\mbox{$\Ng \lesssim 1$} for both the GD and the GB.
For the low birth-frequency population of \mbox{$\pellip{-8}{-7}$}, 3G detectors
are needed to reach a constraint of $\Ng \lesssim 100$.
Finally,~\mbox{$\pellip{-9}{-8}$} can only be constrained for high birth-frequencies 
to \mbox{$\Ng \lesssim 100$} using 3G detectors.

Our results suggest the all-sky searches using the current generation of ground-based
interferometric detectors will not be able to constrain the \mbox{$\pellip{-9}{-8}$} population
of gravitars, as the expected detectability distance is below the required $\SI{100}{\parsec}$.
For the low birth-frequency distribution, this statement remains true even for 3G detectors.
If one takes a less conservative approach and uses $d^{10\%}$ instead of $d^{90\%}$
(see Fig.~\ref{fig:distance_plot_low} and Fig.~\ref{fig:distance_plot_high}),
this population can be meaningfully constrained in O4 or O5. As discussed in
Sec.~\ref{subsec:population}, the distribution of these sources is highly uncertain,
as they closely follow their (highly uncertain) birth frequency distribution. 
The success of searches such as~\cite{Dergachev:2020fli, Dergachev:2020upb,Dergachev:2021ujz}
is thus contingent on selecting an appropiate parameter space so that a sufficiently high
number of sources is present.

Albeit we focus on blind searches for \emph{isolated} sources, these results can be taken as
an ``optimistic estimate'' for sources in binary systems, as in such cases the sensitivity
is degraded due to extended parameter-space breadth\footnote{This ``optimistic estimate''
becomes accurate if one considers binary systems whose properties
are consistent with an isolated source~\cite{Singh:2019han}.
}~\cite{Tenorio:2021wad, Covas:2019jqa, Covas:2020nwy, Tenorio:2021sfj, Wette:2023dom}.

In particular, Ref.~\cite{Covas:2022rfg} sets upper limits on the ellipticity and r-mode saturation
amplitude of nearby millisecond NSs in binary systems. 
The observed millisecond pulsar population provides mild evidence for a minimum ellipticity
of $\ellip \approx 10^{-9}$ to be sustained by these objects~\cite{Woan:2018tey}.
The ellipticity upper limits reported in~\cite{Covas:2022rfg} are $\ellipul \approx 10^{-8}$,
about an order of magnitude higher, at \SI{10}{\parsec};
the expected number of NS at such distance,
as shown in Fig.~\ref{fig:number_gravitars}, \mbox{is $\lesssim 1$}.
The r-mode-amplitude upper limits at \SI{100}{\parsec}, where about 100 NSs are expected,
start to dive below the upper end of theoretical estimates; the lower end, however,
is about two orders of magnitude lower.

Our results suggest an improvement in sensitivity of an order of magnitude (e.~g.~3G detectors)
\emph{will not be sufficient} to reach $\ellipul \approx 10^{-9}$ at the required distance
to constrain a significant fraction of the population \emph{using all-sky searches}.
For the case of r-modes, Ref.~\cite{Maccarone:2022bym} proposes the existence of a substantial
population of undetected quiescent low-mass X-ray binaries within \SI{1}{\kilo\parsec}.
 Using a similar argument as for the case of isolated NSs,
\emph{all-sky searches} are unlikely to constrain the population of binary sub-kiloparsec
r-mode sources across most of the theoretically expected r-mode amplitudes.

\subsection{Alternative blind-search strategies}
\label{sec:less_breadth}

Reducing the breadth of a search tends to increase the resulting sensitivity due to a reduction of the 
``trials factor''~\cite{Wette:2011eu,Dreissigacker:2018afk,Tenorio:2021wad,Wette:2023dom}.
Increasing the sensitivity depth of a search translates directly into an increase
on the detectability distances listed in Table~\ref{table:distance_90}. 
This is mainly caused by two mechanisms.

First, given a computing budget, narrower searches can afford to use more sensitive methods,
which are less likely to produce highly-significant noise candidates.
As a result, the sensitivity of a search increases at a given
false-alarm probability~\cite{Wette:2011eu,Wette:2018bhc}.
These methods,  however, tend to be less robust to unaccounted
physics~\cite{Ashton:2017wui,Mukherjee:2017qme}, which makes them undesirable 
to conduct blind searches.

Second, most all-sky searches follow a hierarchical
scheme~\cite{Brady:1998nj,Krishnan:2004sv}: the main search stage
returns a certain number of candidates to be followed-up using
alternative methods~\cite{Papa:2016cwb,2020ApJ...897...22P, Tenorio:2021njf}.
The computing cost of typical follow-up methods is negligible with respect to the search's.
Reducing the analyzed parameter-space while maintaining the number of candidates
to follow-up increases the false-alarm probability of the main search stage.
This makes weaker signals more likely to be detected by a follow-up method,
and thus increases the overall detection probability of the search.

Narrowing the parameter space, on the other hand, also reduces the number of
potential sources probed by a search; this may end up counteracting the sensitivity
gains if an unfavourable parameter-space region is chosen.

Estimating the increase in sensitivity depth due to a reduction of
the analyzed parameter space is a complex topic~\cite{Dreissigacker:2018afk,Wette:2023dom}.
In App.~\ref{app:breadth}, we empirically compute the increase in sensitivity depth
for comparable search pipelines targeting different parameter-space regions.
A breadth reduction of an order of magnitude, such as the
result of using the astronomical priors proposed in~\cite{2014AN....335..935S},
tends to increase sensitivity depth by less than 20\%.
Substituting an all-sky search with a \emph{spotlight search},
which covers a sky are three orders of magntiude smaller,
tends to increase the sensitivity depth by less than a factor 2. 
Note that these improvements are comparable to the expected sensitivity
improvement from O3 to O4.

Let us assume an optimistic scenario where we select a favourable parameter-space region
such that the sensitivity depth increases by a factor 2 (i.e.~the maximal increase 
discussed in App.~\ref{app:breadth}). Under this assumption, we can constrain a population
of gravitars down to \mbox{$\Ng \lesssim 100$} whenever
\mbox{$d^{90\%} \gtrsim \SI{50}{\parsec}$} in Tabel~\ref{table:distance_90}.
This implies that such a constraint would be imposed on the low birth-frequency
\mbox{$\pellip{-7}{-6}$} population and the high birth-frequency \mbox{$\pellip{-8}{-7}$} 
population using O3 rather than O4 data. The constraints for the \mbox{$\pellip{-9}{-8}$}
population \emph{remain unchanged despite the significant increase in sensitivity};
in other words, \emph{we do not expect alternative blind-search strategies
to constrain a broader class of CW-source populations}.

On the other hand,  reducing the parameter-space breadth of a search reduces the 
search's computing cost by about the same factor,  as computing cost is approximately proportional
to the number of templates considered in a search. For the specific case of~\cite{2014AN....335..935S},
this implies between 5 to 10 times less computing cost than an all-sky search with a slight increase
in sensitivity.

\section{Conclusion}
\label{sec:conclusion}
We have discussed the capability of blind searches to constrain the sub-kiloparsec
population of young CW sources. Our results suggest all-sky searches using Advanced LIGO
detectors will be able to place astrophysically meaningful constraints on gravitars with
ellipticities greater than \mbox{$\ellip \approx 10^{-7}$} regardless of their birth frequency.
Once 3G detectors become operational, these constraints will be extended to
\mbox{$\pellip{-8}{-7}$} for all birth frequencies and \mbox{$\pellip{-9}{-8}$} for
birth frequencies  above \SI{1}{\kilo\hertz}. We forsee no astrophysically meaningful
constraints will be set for the sub-kiloparsec population of gravitars with ellipticities lower than
$\ellip \approx 10^{-8}$ born below \SI{1}{\kilo\hertz}. These results put into question
whether all-sky searches are the most appropriate tool to study the sub-kiloparsec
population of unknown weakly-emitting CW sources~\cite{Dergachev:2020fli,
Dergachev:2020upb, Dergachev:2021ujz,Covas:2022rfg}.

We have also explored the effect of using astronomical priors to guide all-sky
searches~\cite{2014AN....335..935S,LIGOScientific:2015ina,Dergachev:2019pgs}.
Our results suggest these strategies will not be able to place significantly different
constraints with respect to all-sky searches. The computing cost of the resulting searches,
however, can potentially be reduced by several orders of magnitude, as it scales almost
linearly with parameter-space breadth.

Additionally, the gravitar populations described in this work can be used to setup future all-sky
searches for CW from isolated objects. For example, the ellipticity upper limits at \SI{1}{\kilo\parsec}
in Fig.~\ref{fig:O3_ul} still allow for a population of $\pellip{-6}{-5}$ gravitars to be present below
\SI{250}{\hertz}. This frequency range is precisely where a $\pellip{-6}{-5}$ population of young
gravitars is expected to be found, and is located near the most sensitive band of the Advanced LIGO
interferometric detectors (see Fig.~\ref{fig:asd}). Given that computing cost scales quadratically with
frequency~\cite{Prix:2006wm}, this parameter space provides a favorable detection probability
versus computing cost trade-off with respect to broader searches such
as~\cite{KAGRA:2022dwb,Steltner:2023cfk,Dergachev:2022lnt}.

Finally, we note the conservative approach taken in this work. The population of sub-kiloparsec
NSs is expected to contain about 100 objects~\cite{Popov:2003iq,Popov:2003hq}, out of which
an unknown fraction would behave akin to a gravitar; as a result, we chose $d^{90\%}$ to account for
the likely small number of sources. The case of claiming a first CW detection, 
on the other hand, corresponds to detecting a smaller fraction of the population.
For instance, the estimated distances in Table~\ref{table:searches} increase
by an order of magnitude for $d^{10\%}$
(see Fig.~\ref{fig:distance_plot_low} and Fig.~\ref{fig:distance_plot_high}).
This makes low-ellipticity sources detectable by current
searches~\cite{Dergachev:2020fli, Dergachev:2020upb, Dergachev:2021ujz,Covas:2022rfg}.
With this, the sub-kiloparsec population of unknown sources is indeed a good candidate to
provide a first CW signal detection.

\section*{Acknowledgements}
I thank 
Alicia M. Sintes, 
David Keitel,
Aditya Sharma,
Karl Wette,
Joan-René Mérou,
Rafel Jaume,
and Miquel Oliver
for fruitful discussions.
I thank Joe Bayley, 
and the CW working group of
the LIGO-Virgo-KAGRA Collaboration
for comments on the manuscript.
This work was supported by the Universitat de les Illes Balears (UIB);
the Spanish Agencia Estatal de Investigaci\'on grants PID2022-138626NB-I00,
PID2019-106416GB-I00, RED2022-134204-E, RED2022-134411-T,
funded by MCIN/AEI/10.13039/501100011033;
the MCIN with funding from the European Union NextGenerationEU/PRTR (PRTR-C17.I1);
Comunitat Aut\`onoma de les Illes Balears through the Direcci\'o General de Recerca,
Innovaci\'o I Transformaci\'o Digital with funds from
the Tourist Stay Tax Law(PDR2020/11 - ITS2017-006),
the Conselleria d’Economia, Hisenda i Innovaci\'o grant numbers
SINCO2022/18146 and SINCO2022/6719,
co-financed by the European Union and FEDER Operational Program 2021-2027
of the Balearic Islands; the ``ERDF A way of making Europe'';
and EU COST Action CA18108.
This material is based upon work supported by NSF's LIGO Laboratory
which is a major facility fully funded by the National Science Foundation.
This document has been assigned document number LIGO-P2300346.

\bibliographystyle{apsrev4-2}
\bibliography{references}

\appendix

\section{Sensitivity impact due to narrower searches}
\label{app:breadth}
Modelling the sensitivity of arbitrary realistic search setups is complicated
due to the complexity of available configuration
choices~\cite{Dreissigacker:2018afk, Wette:2023dom}.
Instead, we select searches using comparable methods covering different parameter-space regions 
in a given observing run and compare their $h_0$ upper limits along overlapping frequency bands.
The result is a broad estimate of the expected increase in sensitivity depth
due to a parameter-space reduction. 
This approach cannot distinguish between the two effects discussed in Sec.~\ref{sec:less_breadth},
but the results here discussed will provide an educated estimate on the potential sensitivity
improvement stemming from a reduction of the search's parameter-space breadth. 

We start by comparing three all-sky searches in O2 data. Specifically, we take the
\texttt{Einstein@Home} search ~\cite{Steltner:2020hfd} as a base case and compare
the sensitivity to the \texttt{Falcon} low-ellipticity
search~\cite{Dergachev:2020fli, Dergachev:2020upb, Dergachev:2021ujz}
and the \texttt{Weave} deep exploration search~\cite{Wette:2021tbv}.
Both the \texttt{Falcon} and \texttt{Weave} searches cover a limited spindown range,
about 4 to 5 orders of magnitude narrower than the \texttt{Einstein@Home search}.
The resulting sensitivity depth is 30\% higher for \texttt{Weave}
and 50\% higher for \texttt{Falcon}. This discrepancy of about a factor 2 is due to the lack
of follow-up stages in the \texttt{Weave} search compared to the \texttt{Falcon} search.  

We then compare the O3a all-sky \texttt{Einstein@Home} search~\cite{Steltner:2023cfk}
to the O3a all-sky \texttt{Falcon} search~\cite{Dergachev:2022lnt}.
The spindown range is about 2 orders of magnitude narrower.
In this case, depth increase is limited to a $5\%$.

Finally, we compare the sensitivity depth increase of \emph{spotlight} searches,
which aim at a reduced sky area, with respect to all-sky searches. Specifically,
we compare the S6 \texttt{Powerflux} spotlight search~\cite{LIGOScientific:2015ina}
to the S6 all-sky search results reported by \texttt{Powerflux}~\cite{LIGOScientific:2016bah},
and the O1 spotlight loosely coherent search~\cite{Dergachev:2019pgs} to the
O1 all-sky search results reported by \texttt{Powerflux}~\cite{LIGOScientific:2018gpj}.
These spotlight searches focus on a disk in the sky with a radius of \SI{0.06}{\radian},
which corresponds to a 3 order of magnitude reduction with respect to an all-sky search.
The sensitivity depth in crease is about \mbox{$(60 - 80)\%$} for the S6 searches and
\mbox{$(80 - 100)\%$} for the O1 searches. The slightly bigger increase in O1 results may be
attributed to the use of a more sensitive method for the spotlight search (``loose coherence'')
than for the all-sky search (\texttt{Powerflux}).

For completeness, we simulate the increase in sensitivity depth due to a reduction of 
an order of magnitude in sky area by increasing the false-alarm probability by 
one to two orders of magnitude $p_{\mathrm{fa}}$
in~\texttt{octapps}~\cite{Dreissigacker:2018afk}
using the \texttt{Weave} setup proposed in~\cite{2019PhRvD..99h2004W}.
Sensitivity depth improvements are on the order of $10\%$ to $20\%$, depending on the
initial false alarm probability of the search.

\end{document}